# Professional Insights into Benefits and Limitations of Implementing MLOps Principles

Gabriel Araujo[1], Marcos Kalinowski[1], Markus Endler[1], and Fabio Calefato[2]

[1]*Departamento de Informática, Pontifícia Universidade Católica do Rio de Janeiro (PUC-Rio), Brazil*
[2]*Dipartimento di Informatica, Universita` degli Studi di Bari Aldo Moro (Uniba), Italy*
{gcarvalho, endler, kalinowski}@inf.puc-rio.br, fabio.calefato@uniba.it

Abstract: **Context:** Machine Learning Operations (MLOps) has emerged as a set of practices that combines development, testing, and operations to deploy and maintain machine learning applications. **Objective:** In this paper, we assess the benefits and limitations of using the MLOps principles in online supervised learning. **Method:** We conducted two focus group sessions on the benefits and limitations of applying MLOps principles for online machine learning applications with six experienced machine learning developers. **Results:** The focus group revealed that machine learning developers see many benefits of using MLOps principles but also that these do not apply to all the projects they worked on. According to experts, this investment tends to pay off for larger applications with continuous deployment that require well-prepared automated processes. However, for initial versions of machine learning applications, the effort taken to implement the principles could enlarge the project's scope and increase the time needed to deploy a first version to production. The discussion brought up that most of the benefits are related to avoiding error-prone manual steps, enabling to restore the application to a previous state, and having a robust continuous automated deployment pipeline. **Conclusions:** It is important to balance the trade-offs of investing time and effort in implementing the MLOps principles considering the scope and needs of the project, favoring such investments for larger applications with continuous model deployment requirements.

## 1 INTRODUCTION

Machine Learning (ML) is a discipline that allows machines to automatically learn from data and past processing experiences to identify data patterns, classify data, and predict results with minimal human intervention (Jordan and Mitchell, 2015). Supervised ML is a sub-branch of ML (Shetty et al., 2022) that depends on a human domain expert who 'teaches' the learning scheme with the required supervision, typically by annotation/labeling data mapping inputs to selected outputs. Supervised ML problems are grouped into classification and regression. In classification problems, the prediction results correspond to discrete values. In regression, on the other hand, the results correspond to continuous values. A review of ML algorithms can be found in (Ray, 2019).

Online ML refers to applications of ML where data becomes available in sequential order and is used to update the best predictor for future data at each step, as opposed to batch learning techniques which generate the best predictor by learning on the entire training data set at once. The goal of online learning is to make a sequence of accurate predictions given the knowledge of the correct answer to previous prediction tasks and possibly additional available information (Shalev-Shwartz, 2012). It is commonly used in situations where it is necessary for the algorithm to adapt to new patterns in the data dynamically or when the data itself is generated as a function of time, for example, stock price prediction.

Unfortunately, the success of many real-world ML applications falls short of expectations (Kocielnik et al., 2019). Many ML projects fail and never reach production (van der Meulen and McCall, 2018). From a research perspective, this does not come as a surprise, as the ML community has focused extensively on the building of ML models but not on (a) building production-ready ML products and (b) providing the necessary coordination of the resulting, often complex, ML system components and infrastructure, including the roles required to automate and operate an ML system in a real-world setting (Posoldova, 2020). For example, in many industrial applications, data scientists still manually manage ML workflows to a great extent, resulting in many problems during the operations of the respective ML solution (Lwakatare et al., 2020). This is a particular problem in online learning, where data arrives in sequential order, and the model is expected to learn and update the best predictor for future data at every step.

To help solve problems such as building production-ready applications for complex systems, the technical community has started to adopt continuous software engineering practices such as Development and Operations (DevOps). DevOps can be defined as the development method emphasizing software delivery, automated deployment, continuous integration, and quality assurance (Jabbari et al., 2016). The practice of continuous delivery of ML solutions is called Machine Learning Operations (MLOps), which mimics DevOps practices but introduces additional actions specific to ML (Mäkinen et al., 2021).

MLOps is a paradigm including best practices and sets of concepts and a development culture regarding the end-to-end conceptualization, implementation, monitoring, deployment, and scalability of ML products (Kreuzberger et al., 2023). MLOps aims to bridge the gap between development and operations for ML-enabled systems and represents the alignment between the building of ML models, software development, and operation (Kalinowski et al., 2023).

Academic research has focused intensively on the building and benchmarking of ML models but little on the operation of complex ML systems in real-world scenarios. In the real world, adopting software engineering best practices in MLOps is still limited (Kreuzberger et al., 2023).

The goal of this paper is to gather practitioner insight into the benefits and limitations of using MLOps principles in the context of online supervised learning. To this end, we conducted two focus group sessions with six experienced ML developers. The focus group revealed key benefits (*e.g.*, reducing manual errors, facilitating rollback, establishing an automated model deployment pipeline) but also found that investing effort in its principles tends to be more rewarding for larger, continuously deployed applications needing automated processes. For initial ML application versions, implementing MLOps principles can expand the project scope and delay production deployment.

## 2 MLOPS

Machine Learning Operations (MLOps) is a core function of ML engineering, focused on streamlining the process of deploying ML models to production and then maintaining, scaling, and monitoring them. MLOps is a collaborative endeavor that often involves data scientists, DevOps engineers, and IT. An optimal MLOps experience is one in which ML assets are treated consistently with all other software assets within a CI/CD environment (Kalinowski et al., 2023). *I.e.*, ML models can be deployed alongside the services that wrap them and the services that consume them as part of a unified release process (Visengeriyeva et al., 2023).

As ML is increasingly pervasive in software products, we need to establish best practices and tools to test, deploy, manage, and monitor ML models in real-world production. In short, with MLOps, we strive to avoid 'technical debt' in ML applications. Hereafter, we briefly describe the MLOps principles that provide the background for this paper: Automation, Monitoring, Versioning, Reproducibility, Testing, and Deployment (Visengeriyeva et al., 2023).

**Automation:** The objective of an MLOps team is to automate the deployment of ML models in the core software system or as a service component. This means automating the end-to-end ML workflow pipeline without any manual intervention.

**Monitoring:** After the ML model has been deployed, the model monitoring step aims to monitor it to ensure that the ML model performs as expected. Monitoring an ML application is important to understand problems with the data, model, and application.

**Versioning:** The MLOps versioning principle consists of organizing and versioning code, datasets, and models. In an ML project, data scientists continuously work on developing new models. This process relies on trying different combinations of data, parameters, and algorithms.

**Reproducibility:** The reproducibility principle is described as the process of repeatedly running an ML application on certain datasets and obtaining the same or similar results. The reproducibility principle ensures that researchers can reproduce the accuracy of reported results and detect biases in the models.

**Testing:** Testing and monitoring are important strategies for improving reliability, reducing technical debt, and reducing long-term maintenance costs. The MLOps testing principle introduces tests for features and data, model development, and ML infrastructure as part of ensuring data and model quality.

**Deployment:** The deployment of ML models or pipelines is the process of making models available in production environments so that applications and APIs can consume the trained model. The MLOps deployment principle consists of containerizing the ML stack and providing access to the deployed model.

# 3 FOCUS GROUP DESIGN

To assess the MLOps principles from the practitioners' point of view, we designed a focus group to promote in-depth expert discussions about the benefits and limitations of applying these principles within ML projects. Focus group is a qualitative research method based on collecting data through group sessions, which allow the extraction of experiences from participants (Kontio et al., 2008). A focus group session is planned to address in-depth discussions about a particular topic during a controlled time slot. Focus group studies have been conducted in software engineering to reveal consolidated expert insights and feedback (*e.g.* (Martakis and Daneva, 2013), (Almeida et al., 2023)). We decided to use a focus group as a suitable option to understand practitioners' perceptions of the benefits of MLOps for supervised online ML applications.

We conducted two focus group sessions with six expert ML developers (three in each session) who have experience creating large-scale ML applications and have had contact with both MLOps- and non-MLOps-based ML application development.

## 3.1 Context and Participant Characterization

We selected participants from three different organizations to gain insight from various perspectives. The characterization of the participants is shown in Table 1. We observe that all the participants have a high level of knowledge and at least 3 years of experience developing machine-learning applications. Despite their expertise and having worked on MLOps-based projects, participants did not consider themselves highly knowledgeable about MLOps. This could be because many companies seem to be still maturing their MLOps approaches, with data scientists still manually managing ML workflows to a great extent (Kreuzberger et al., 2023).

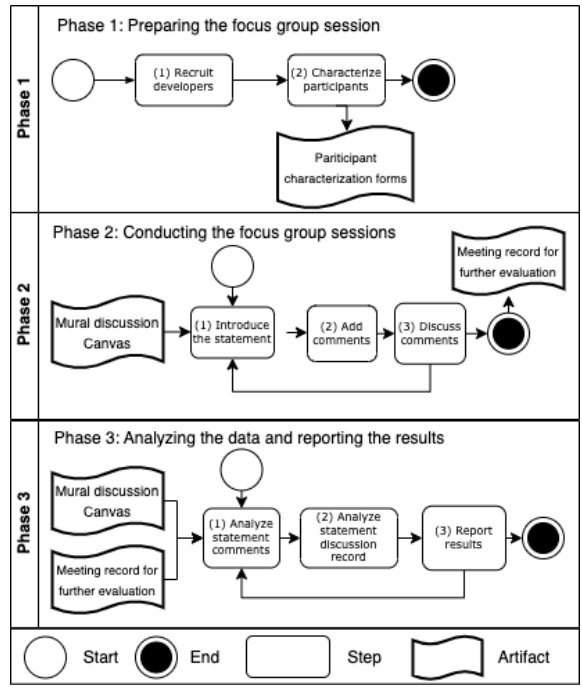

Figure 1: Focus Group Overview.

## 3.2 Focus Group Planning and Design

We carefully designed our focus group following the guidelines proposed by (Kontio et al., 2008). The goal of our focus group can be described following the Goal-Question-Metric goal definition template (Basili and Rombach, 1988), as follows: *Analyze* the MLOps principles *with the purpose of* characterizing *with respect to* the benefits and limitations of the MLOps principles *from the point of view of* ML experts *in the context of* supervised online ML applications.

Figure 1 shows the steps adopted throughout the focus group. We organized these steps into three major phases: **(1)** Preparing the focus group session; **(2)** Conducting the focus group sessions; and **(3)** Analyzing the data and reporting the results. In the following, we describe each phase and step.

**Phase 1: Preparation for the focus group session.** This phase consists of collecting preliminary resources to support the execution of the focus group session. For this purpose, we follow two steps.

**Step 1: Recruit developers** consisted of recruiting developers with experience in ML and MLOps to participate in discussions. We contacted developers from three different organizations and industries to participate in our study. We obtained the acceptance of six experts using a consent form in which we explained our research goals and that the information provided by each participant would be treated confi-dentially and used for study purposes only.

Table 1: Focus group 1 (P1, P2, and P3) and focus group 2 (P4, P5, and P6) participants information.

| ID | Graduation Level | Years of Experience with ML | Classification of knowledge in ML | Classification of knowledge in MLOps | Job Title | Industry | Company Size |
|---|---|---|---|---|---|---|---|
| P1 | Bachelor degree | 3 | High | Medium | ML Engineer | Retail and e-commerce | 50,000+ |
| P2 | Masters Degree | 4 | High | Medium | Data Scientist | Oil & gas | 40,000+ |
| P3 | Masters degree | 5 | High | Medium | Data Scientist | Oil & gas | 40,000+ |
| P4 | Masters degree | 3 | High | Medium | ML Engineer | Oil & gas | 40,000+ |
| P5 | Masters degree | 4 | High | Medium | ML Engineer | Oil & gas | 40,000+ |
| P6 | Bachelor degree | 5 | Very High | High | DS Specialist | Finance | 3,000+ |

Next, **Step 2: Characterize participants** aimed to collect basic information to characterize participants via the Participant Characterization Form. Our main goal was to profile each participant so that we could better interpret our study results. The form asked participants about their academic degree, the number of years working in the field of ML, and a rating of their knowledge of ML and MLOps, which could be evaluated in the following possibilities: very low, low, medium, high, very high (*cf.* Table 1).

**Phase 2: Conducting the focus group sessions.** This phase consists of collecting data on the participants' perception of the benefits and limitations of using the MLOps principles for supervised online ML applications. To discuss the benefits and limitations of MLOps, we derived statements from commonly used ML-based software delivery metrics (deployment frequency, lead-time for change, and mean time to restore) (Visengeriyeva et al., 2023), and asked participants to discuss the effects of using MLOps on these metrics. As these metrics mainly concern automation (including deployment, reproducibility, and testing aspects), we added two additional statements to allow discussion regarding the monitoring and versioning principles. Finally, we added a generic statement on the use of MLOps principles to gather any additional insights that experts would like to provide.

We used an online environment to promote discussions on the benefits and limitations of MLOps principles for ML applications. We designed a template using the MIRO online collaborative platform (Miro, 2023). In practice, using this tool, we were able to build an interactive mural to facilitate the conduction of the focus group sessions.

Our template is divided into 5 columns that seek to understand whether, for each line containing a statement, the participants, through virtual post-its: strongly agree, partially agree, partially disagree, strongly disagree, or have no opinion. Each statement is discussed in isolation, and we defined the dynamics of the focus group session in three steps as follows.

**Step 1: Introduce the statement** aimed to present each statement. For this purpose, the moderator of the session read this information out loud. Next, in **Step 2: Add comments**, we asked each participant to add one or more *post-its* for each comment they had on the statement, placing them in the appropriate columns reflecting their agreement. Finally, in **Step 3: Discuss comments**, we asked participants to explain their comments (and why they agreed or disagreed with the statement) and discussed them within the group. Each comment should be documented on the post-it, as shown in Figure 2. The comment was just a brief summary of the reason they selected a column, and we constantly asked participants to share the knowledge and experiences surrounding the statement to enrich the discussions and understanding. Additionally, whenever the moderator felt that a comment was poorly written, he asked the participants to provide further considerations.

**Phase 3: Analyzing the data and reporting the results.** The focus group sessions were conducted online via Zoom. Additionally, we kept video and audio records of the sessions to support the data analysis. Both sessions were held in August 2023. We analyzed the comments for each statement, referring to the transcribed audio discussion records to better understand what the developers meant with each note. The audio records of the Zoom recording were transcribed using a popular speech-to-text transcription model (Grosman, 2021). We report the results of the discussions for each statement in the following section.

## 4 RESULTS: PRACTITIONER INSIGHTS

We asked participants to choose and justify whether they agreed or disagreed with the statements reported in the context of the benefits and limitations of using the MLOps principles for supervised online ML applications. We have collected these comments through post-it notes added by participants in the session's MIRO board (as illustrated in Figure 2). Each participant had a specific color of post-it and could add more than one comment.

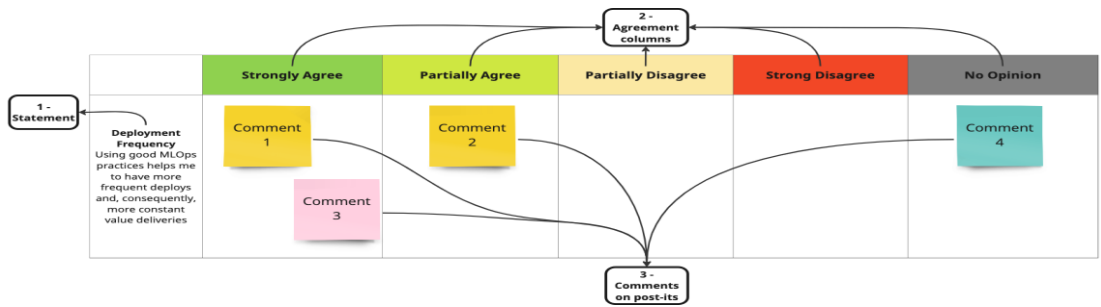

Figure 2: Focus Group Template Steps.

In order to analyze these comments, we first watched the video and automatically transcribed its audio into plain text. Then, we analyzed all post-it comments written by the participants and associated transcription quotes. In the following subsections, we summarize the agreement for each statement, also qualitatively providing relevant comments that emerged during the discussions. A full qualitative analysis of the discussion with all the details of the sentences uttered during the discussions can be found in our online repository (Araujo et al., 2023). Complete recordings and transcriptions were not made available to preserve anonymity.

**Statement 1 - Deployment frequency: Using MLOps principles helps me to have more frequent deployments and, consequently, more constant value deliveries.** Figure 3 summarizes the agreement positioning of the comments by the participants for this statement. Due to space constraints and for the sake of readability, the detailed post-it comments (represented by an 'X' in Figure 3) are provided in our online repository, while herein we summarize the main insights taken from the discussions.

|  | P1 | P2 | P3 | P4 | P5 | P6 |
|---|---|---|---|---|---|---|
| Strongly Agree | X | X | X |  | X | X |
| Partially Agree |  |  |  | X | X |  |
| Partially Disagree |  |  |  |  |  |  |
| Strongly Disagree |  |  |  |  |  |  |
| No Opinion |  |  |  |  |  |  |

Figure 3: Statement 1 focus group opinion

We observe that four participants provided comments strongly agreeing; one participant reported a comment related to strong agreement and another comment related to partial agreement; and one participant reported a comment regarding partial agreement. Therefore, in terms of deployment frequency, participants mainly agree that the MLOps principles help generate more frequent deployments and value deliveries. During the discussion, participant P3 mentioned difficulties related to pipeline automation "*If a pipeline is robust with good practices, the value ends up being very high, but getting there is complicated.*"

In light of such difficulties, participants also mentioned that if a product doesn't require a high deployment frequency, it might not be worth the effort to prepare a well-structured pipeline following the MLOps principles, which requires development and preparation time, instead of using that effort on other fronts. For example, P6 argued that "*If you have a far away deadline and a well-defined delivery roadmap, the impact of MLOps ends up being huge. However, if the first version of the product delivers what you need in practice, MLOps ends up having a low impact.*"

**Statement 2 - Lead time for changes: Using MLOps principles helps me reduce the time for delivery and deployment, counting from the moment the code is merged.** This statement was intended to collect the experts' experiences on how the automation principle can affect the lead time for changes. *i.e.*, do the MLOps principles help the developers to have code merged faster into a production? As shown in Figure 4, four participants reported strongly agreeing comments, another participant reported one strongly agreeing and one partially agreeing comment, and one reported a partially agreeing comment.

|  | P1 | P2 | P3 | P4 | P5 | P6 |
|---|---|---|---|---|---|---|
| Strongly Agree | X |  | X | X | X | X |
| Partially Agree | X | X |  |  |  |  |
| Partially Disagree |  |  |  |  |  |  |
| Strongly Disagree |  |  |  |  |  |  |
| No Opinion |  |  |  |  |  |  |

Figure 4: Statement 2 focus group opinion

Hence, for lead time to production, it is also possible to observe an overall agreement on the reduc-tion of the time for delivery and deployment. Participant P6 emphasized that "*practices such as versioning and automation help corrections as a matter of urgency. When you incorporate these practices, you reduce manual steps and human errors, standardizing changes and speeding up delivery.*" However, a similar counterpoint seen in the previous statement was raised by participant P1, who mentioned that achieving pipeline maturity is time-consuming and, consequently, requires effort and time.

**Statement 3 - Mean Time to Restore: From the moment an incident occurs and there is a need for rollback, I can easily go back to my model in the previous version, without using continuous deployment practices.** This statement was set to understand how the MLOps principles might influence moments of urgency, such as a bug in a production environment. It addresses the mean time to restore metric, *i.e.*, the time it takes an application to recover, usually through a rollback, to a functional state from a non-functional state. This process tends to be done at moments of tension and stress, as a problem in production can cause damage to the application model, ruining data, as well as possible financial damage to the company, depending on what the application is used for. Note that the statement intentionally aimed to evaluate the agreement on the feasibility of restoring without continuous deployment practices. Our goal was to understand whether, in the experience of our participants, this process done manually and without reproducibility principles and continuous deployment practices is considered feasible.

|  | P1 | P2 | P3 | P4 | P5 | P6 |
|---|---|---|---|---|---|---|
| Strongly Agree |  |  |  |  |  |  |
| Partially Agree |  |  |  |  |  |  |
| Partially Disagree | X |  |  |  | X | X |
| Strongly Disagree |  | X | X | X |  |  |
| No Opinion |  |  |  |  |  |  |

Figure 5: Statement 3 focus group opinion

As shown in Figure 5, three participants reported partially disagreeing, and three reported strongly disagreeing. Therefore, based on participants' experience, an ML application with a structured pipeline that follows the principles of automation and versioning tends to make the reproducibility and rollback process easier and more successful, avoiding error-prone manual steps. Participant P4 raised an interesting point of view: the human side of having a problem in production. According to the participant, "*There is an impact on people under pressure during a crisis, such as a bug in production, increasing the probability of human error. Therefore, an automated continuous deployment process is one of the main benefits of following the MLOps principles.*"

**Statement 4 - Monitoring: Setting up alarms and monitoring can be easily done without the use of MLOps.** This statement's goal is to evaluate the use of MLOps to ensure monitoring by reflecting on whether, based on experts' experience, it was possible to achieve monitoring and alarm quality in their projects without relying on the MLOps principles. As can be seen in Figure 6, one participant reported partially agreeing, two partially disagreeing, and one strongly disagreeing.

|  | P1 | P2 | P3 | P4 | P5 | P6 |
|---|---|---|---|---|---|---|
| Strongly Agree |  |  |  |  |  |  |
| Partially Agree |  |  |  |  |  | X |
| Partially Disagree |  |  |  | X | X |  |
| Strongly Disagree |  |  | X |  |  |  |
| No Opinion | X | X |  |  |  |  |

Figure 6: Statement 4 focus group opinion

Although the participants' opinions about the statement were diverse, a common aspect was that monitoring is usually left behind from the application scope in the early stages. P6 noted that in their experience in different projects and companies, the monitoring part took a back seat during the prioritization of the project tasks. He mentioned that "*commonly, the implementation of monitoring and alarms arise when the first pain of not having monitoring occurs. In these cases, the negative experience generated by the lack of monitoring creates an urgency to develop it and, consequently, in the need to have something implemented quickly, the practices presented in the MLOps principle were left behind.*" In line with this, participant P5 mentioned that "*even in projects I've worked on that didn't use MLOps, we already had some alarms that worked well.*"

Participant P3 emphasized that "*although it is possible to set up monitoring and alarms without MLOps, it improves speed, agility, automation, and integration for more effective and consistent monitoring practices. Creating alarms and monitoring without using MLOps is like having a hammer and missing the other tools needed to build a house.*" The discussion allowed us to conclude that it is possible to build monitoring without following the MLOps principles; however, implementing the MLOps principles will facilitate and come into great hands when developing alarms and monitors.

**Statement 5 - Versioning: In my projects that did not use the MLOps principles, I was able to migrate between the deployments made from my service.** Through this statement, our goal is to understand from the opinions of experts whether they consider it feasible to carry out common actions such as rollback and model version exchange (reproducibility) in projects that did not use the MLOps principles. As shown in Figure 7, two participants reported partially disagreeing, and four reported strongly disagreeing with the statement.

|  | P1 | P2 | P3 | P4 | P5 | P6 |
|---|---|---|---|---|---|---|
| Strongly Agree |  |  |  |  |  |  |
| Partially Agree |  |  |  |  |  |  |
| Partially Disagree |  |  | X |  | X |  |
| Strongly Disagree | X | X |  | X |  | X |
| No Opinion |  |  |  |  |  |  |

Figure 7: Statement 5 focus group opinion

We observe an overall disagreement on the feasibility of developing a complex production-ready ML application without effective versioning. A common point in the comments is that the versioning principle is essential for achieving quality. Versioning is important in making reproducibility possible by enabling previous datasets and models to be identified and facilitating the application to roll back to previously trained models and labeled datasets. Participant P2 stated, "*the effort spent in the short term is rewarded exponentially over a short period of time through the common need to migrate between versioning code, data, and models. You gain little by not having, and you lose a lot by not having.*"

**Statement 6 - General MLOps principles: I understand that the MLOps principles help me deliver faster and, more concisely, generate greater value for my application.** This statement was added to gather additional information that experts would like to provide on the use of the MLOps principles. Therefore, the discussion of this last topic was almost a summary of all the points covered in the previous statements. As can be seen in Figure 8, we had five participants who reported strongly agreeing and one who reported partially disagreeing.

Although there is mainly agreement, we observe that the use of MLOps principles is closely related to the need and capacity to implement them. Participant P1 mentioned: "*You need to know your reality and your problem in order to know how to act on it. MLOps is not a job, it's a culture, and if you want to use MLOps you have to face it as a culture and bring

|  | P1 | P2 | P3 | P4 | P5 | P6 |
|---|---|---|---|---|---|---|
| Strongly Agree | X | X | X |  | X | X |
| Partially Agree |  |  |  |  |  |  |
| Partially Disagree |  |  |  | X |  |  |
| Strongly Disagree |  |  |  |  |  |  |
| No Opinion |  |  |  |  |  |  |

Figure 8: Statement 6 focus groups opinion

*people into your team who follow and believe in that culture to get there.*" Despite the surprising opposite positioning of participant P4, the explanation of why the participant chose to disagree was considered fair and reasonable by the other participants. In addition to mentioning culture, he pointed out that "*the problem is not the characteristics of MLOps but rather the academic and industry environments, which do not have MLOps as part of their culture. As few companies use these principles today, their benefits are not seen competitively. [...] Implementing the MLOps principles could imply changing and delaying the planned roadmap, which is usually focused on delivering usable results.*"

## 5 CONCLUDING REMARKS

This paper assessed the benefits of MLOps for supervised online ML applications. We conducted two focus group sessions with six experienced ML developers. The focus group findings indicate that while ML developers recognize multiple advantages of applying MLOps principles, they also acknowledge that these principles are not universally applicable. MLOps implementation is seen as beneficial primarily for larger projects that involve continuous deployment and need robust automated processes. However, for smaller or initial versions of ML applications, the effort required to adhere to MLOps principles can unnecessarily expand the project's scope and delay the production deployment of the first version. Developers must weigh the time and effort required to implement MLOps against the specific needs and scale of their project.

According to the discussions, the main benefits of adhering to the MLOps principles include enhancing the deployment frequency capacity and shortening the time required to implement changes. It also reduces the risk associated with manual interventions by allowing applications to revert to their previous states and establishing a solid, automated deployment pipeline. Furthermore, implementing MLOps principles can streamline the creation of alarms and monitoring systems and enable systematic version control of code, models, and data.

However, the adoption of MLOps principles within the industry is not yet widespread, and the management of large, complex real-world ML applications lags in terms of scientific investigation. Indeed, while our focus group discussions provide some valuable insights into benefits and limitations of MLOps principles for supervised online ML applications, the findings reflect the experiences of the participants, and other types of empirical studies should be conducted to further assess the effects of applying MLOps principles in different contexts.

# ACKNOWLEDGEMENTS

The authors would like to thank the participants of the focus group sessions, the Brazilian Council for Scientific and Technological Development (CNPq, grant #312275/2023-4), and the Brazilian Higher Education Improvement Coordination (CAPES, finance code 001). The research was also co-funded by projects DARE (code: PNC0000002, CUP: B53C22006420001), SERICS (code: PE0000014, CUP: H93C22000620001), and QualAI (PRIN2022 grant n.2022B3BP5S, CUP: H53D23003510006).

# REFERENCES

Almeida, C., Kalinowski, M., Uchôa, A., and Feijó, B. (2023). Negative effects of gamification in education software: Systematic mapping and practitioner perceptions. *Information and Software Technology*, 156:107142.

Araujo, G., Kalinowski, M., Endler, M., and Calefato, F. (2023). Artifacts: Professional insights into benefits and limitations of implementing mlops principles. https://zenodo.org/doi/10.5281/zenodo.10685254.

Basili, V. R. and Rombach, H. D. (1988). The tame project: Towards improvement-oriented software environments. *IEEE Transactions on software engineering*, 14(6):758–773.

Grosman, J. (2021). Fine-tuned XLSR-53 large model for speech recognition in English. https://huggingface.co/jonatasgrosman/wav2vec2-large-xlsr-53-english.

Jabbari, R., bin Ali, N., Petersen, K., and Tanveer, B. (2016). What is devops? a systematic mapping study on definitions and practices. In *Proceedings of the Scientific Workshop Proceedings of XP2016*, XP '16 Workshops, New York, NY, USA. Association for Computing Machinery.

Jordan, M. I. and Mitchell, T. M. (2015). Machine learning: Trends, perspectives, and prospects. *Science*, 349(6245):255–260.

Kalinowski, M., Escovedo, T., Villamizar, H., and Lopes, H. (2023). *Engenharia de Software para Ciência de Dados: Um guia de boas práticas com ênfase na construção de sistemas de Machine Learning em Python*. Casa do Código.

Kocielnik, R., Amershi, S., and Bennett, P. N. (2019). Will you accept an imperfect ai? exploring designs for adjusting end-user expectations of ai systems. In *Proceedings of the 2019 CHI Conference on Human Factors in Computing Systems*, CHI '19, page 1–14, New York, NY, USA. Association for Computing Machinery.

Kontio, J., Bragge, J., and Lehtola, L. (2008). *The Focus Group Method as an Empirical Tool in Software Engineering*, pages 93–116. Springer London, London.

Kreuzberger, D., Kühl, N., and Hirschl, S. (2023). Machine learning operations (mlops): Overview, definition, and architecture. *IEEE Access*, 11:31866–31879.

Lwakatare, L. E., Crnkovic, I., Rånge, E., and Bosch, J. (2020). From a data science driven process to a continuous delivery process for machine learning systems. In *Product-Focused Software Process Improvement: 21st International Conference, PROFES 2020, Turin, Italy, November 25–27, 2020, Proceedings 21*, pages 185–201. Springer.

Mäkinen, S., Skogström, H., Laaksonen, E., and Mikkonen, T. (2021). Who needs mlops: What data scientists seek to accomplish and how can mlops help? In *2021 IEEE/ACM 1st Workshop on AI Engineering-Software Engineering for AI (WAIN)*, pages 109–112. IEEE.

Martakis, A. and Daneva, M. (2013). Handling requirements dependencies in agile projects: A focus group with agile software development practitioners. volume 2013, pages 1–11.

Miro (2023). Miro online collaborative platform. https://miro.com/. (accessed: 09-20-2023).

Posoldova, A. (2020). Machine learning pipelines: From research to production. *IEEE Potentials*, 39(6):38–42.

Ray, S. (2019). A quick review of machine learning algorithms. In *2019 International conference on machine learning, big data, cloud and parallel computing (COMITCon)*, pages 35–39. IEEE.

Shalev-Shwartz, S. (2012). Online learning and online convex optimization. *Foundations and Trends® in Machine Learning*, 4(2):107–194.

Shetty, S. H., Shetty, S., Singh, C., and Rao, A. (2022). Supervised machine learning: Algorithms and applications. *Fundamentals and Methods of Machine and Deep Learning: Algorithms, Tools and Applications*, pages 1–16.

van der Meulen, R. and McCall, T. (2018). Gartner says nearly half of cios are planning to deploy artificial intelligence. https://www.gartner.com/en/newsroom/press-releases/2018-02-13-gartner-says-nearly-half-of-cios-are-planning-to-deploy-artificial-intelligence. Accessed: 2023-03-16.

Visengeriyeva, L., Kammer, A., Bar, I., Kniesz, A., and Plod, M. (2023). Mlops principles. https://ml-ops.org/content/mlops-principles. (accessed: 03-24-2023).